# Negative Photo Conductivity Triggered with Visible Light in Wide Bandgap Oxide-Based Optoelectronic Crossbar Memristive Array for Photograph Sensing and Neuromorphic Computing Applications

Dayanand Kumar[1#], Hanrui Li[1#], Amit Singh[2], Manoj Kumar Rajbhar[1], Abdul Momin Syed[1], Hoonkyung Lee[3], and Nazek El-Atab[1*]

[1#]*These authors have contributed equally to this work*

ABSTRACT: Photoresponsivity studies of wide-bandgap oxide-based devices have emerged as a vibrant and popular research area. Researchers have explored various material systems in their quest to develop devices capable of responding to illumination. In this study, we engineered a mature wide bandgap oxide-based bilayer heterostructure synaptic memristor to emulate the human brain for applications in neuromorphic computing and photograph sensing. The device exhibits advanced electric and electro-photonic synaptic functions, such as long-term potentiation (LTP), long-term depression (LTD), and paired pulse facilitation (PPF), by applying successive electric and photonic pulses. Moreover, the device exhibits exceptional electrical SET and photonic RESET endurance, maintaining its stability for a minimum of 1200 cycles without any degradation. Density functional theory calculations of the band structures provide insights into the conduction mechanism of the device. Based on this memristor array, we developed an autoencoder and convolutional neural network for noise reduction and image recognition tasks, which achieves a peak signal-to-noise ratio of 562 and high accuracy of 84.23%, while consuming lower energy by four orders of magnitude compared with the Tesla P40 GPU. This groundbreaking research not only opens doors for the integration of our device into image processing but also represents a significant advancement in the realm of in-memory computing and photograph sensing features in a single cell.

**Keywords:** Memristor, Neuromorphic Computing, CNN, Photograph sensing, Electrical SET, photonic RESET,

## Introduction

Recently, memristors have attracted significant attention as artificial synapses for large-scale artificial neural networks. This is primarily owing to their ability to emulate natural synaptic responses, fast writing capabilities, extended retention times, simple architecture, low energy consumption, and compatibility with three dimensional (3D) complementary metal–oxide semiconductor (CMOS) integration [1-6]. The resistive switching (RS) mechanism in memristors relies on the formation and disruption of conducting filaments (CFs) within oxide layers [7-9]. Furthermore, memristors with analog switching characteristics exhibit synaptic behavior when subjected to sequential alternating current (AC) pulses. These devices have garnered attention for application in neuromorphic computing [10-12]. In the human brain, which comprises approximately $10^{12}$ neurons and $10^{15}$ synapses, neurons transmit information to one another via synapses that adaptively regulate the flow of potassium ($K^+$) or calcium ($Ca^{++}$) ions [13]. This phenomenon is similar to the operation of filament-based memristive devices. Over the past decade, researchers have explored transition metal oxides and metal nitrides such as $Al_2O_3$, $HfO_2$, $Ta_2O_5$, $ZrO_2$, TaOx, ZTO, AlN, and SiCN to use their promising memristive capabilities [14-21]. While most studies have focused on memristors with digital switching, owing to their highly stable data retention and swift switching speeds, there is growing interest in analog switching with electrical AC pulses to mimic synaptic functions in neuromorphic computing.

In addition to the electrical phenomenon in memristive devices, a few researchers have shown negative photoconduction or photoinduced RESET (conductive filament rupture) in wide bandgap oxides such as HfOx and $SiO_2$, which is the same as electrical RESET or rupture in memristive devices [22, 23]. By adjusting specific parameters, such as the intensity of light and exposure dose, multistate conductance can be achieved [7, 22]. These achievements have opened a new path for wide-bandgap oxide-based memristors for photographic sensing applications. So far, very few studies have shown the photograph sensing characteristics and nonvolatile memory capabilities together in wide bandgap oxide-based memristors; however, the reported findings are not sufficient and require further enhancement for their applications in the near future [22-29].

In this work, we present the artificial synaptic features, photograph sensing and in-memory functionalities of a memristive device based on an ITO/SiO$_2$/Al$_2$O$_3$/Pt configuration. Al$_2$O$_3$ and SiO$_2$ are promising wideband gap dielectric materials owing to their maturity and CMOS compatibility, in addition to their ability to deposit uniform and conformal ultrathin layers using atomic layer deposition. The device demonstrated excellent stable synaptic features, including LTP/LTD, PPF, and AC endurance for at least $10^7$ cycles without any degradation, and the ability to undergo 188 LTP/LTD cycles repeatedly with 37600 electrical pulses. The first-principles calculation reveals that the SiO$_2$/Al$_2$O$_3$ heterostructure exhibits enhanced conductivity in the presence of an electric field owing to a shift toward more metallic band characteristics. This change promotes the formation of filaments within the resistive switching layers. Furthermore, the device exhibits impressive electrical SET and photonic RESET endurance, maintaining its electrical SET and photonic RESET functionalities for over 1200 cycles without any noticeable deterioration. Based on these attributes, our devices enable the implementation of photographic sensing and neuromorphic applications, including image recognition and denoising. Compared with traditional computing hardware, the neuromorphic computing approach exhibits significantly lower power consumption, exceeding four orders of magnitude reduction compared with the GPU. These impressive features render the device highly suitable for neuromorphic computing and photograph sensing applications. Consequently, we successfully developed an innovative complementary metal-oxide-semiconductor

(CMOS)-compatible memristive device utilizing wide bandgap oxides. In addition to exhibiting notable in-memory capabilities and photograph sensing functionalities, the device also shows excellent efficiency for neuromorphic computing applications, all within a single compact memristive cell.

**Results and discussion**
**Device Fabrication and Electrical Characterization**

In Figure 1 (a), we can observe a representation of the biological neural subsystem system, in which a pre-synaptic neuron is connected to a post-synaptic neuron through synapses located between a pre-synaptic axon terminal and a post-synaptic dendrite terminal. The transfer of information from the pre-synaptic axon terminal to the post-synaptic dendrite terminal is intricately linked to the firing mechanisms of calcium and sodium ions (neurotransmitters). This firing process changes the synaptic weight. Electronic memristors are designed to mimic the functionality of biological synapses. These electronic synapses operate on a principle similar to that of their biological counterparts, in which the quantity of neurotransmitters governs the synaptic weight. In memristors, the conductance between the Top Electrode (TE) and the Bottom Electrode (BE) is regulated by the quantity and distribution of oxygen vacancies within the resistive switching (RS) layers ( depicted in Figure 1 (b), representing an electronic synapse) [29-35].

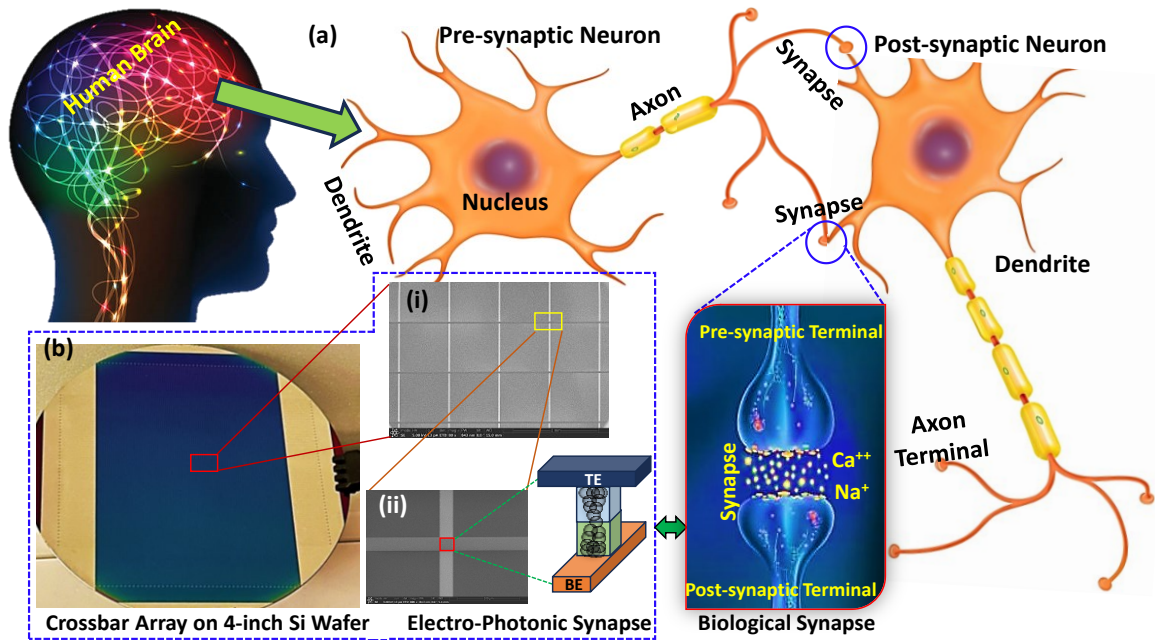

Figure 1. Schematic representation of the biological neural subsystem and an artificial electronic memristive array for a neuromorphic computing system. a) The biological synapse is the junction of axon terminal (per-synaptic neuron) and dendrite terminal (post-synaptic neuron) where the information transmission from pre-synaptic terminal to post-synaptic terminal with the migration of either $K^+$ or $C^{++}$ ions in the human brain. b) The photograph of the 100 × 100 crossbar array, which is fabricated on 4-inch silicon wafer and zoom in SEM images (i) and (ii) with the scale bar: 1 mm and 100 μm, respectively with schematic diagram of ITO/SiO$_2$/Al$_2$O$_3$/Pt crossbar architecture.

Figure 2 presents a comprehensive comparison between previously published wide bandgap oxide-based memristors and the results of our current research. Based on the data presented in Figure 2a and 2b, it is evident that the proposed ITO/SiO$_2$/Al$_2$O$_3$/Pt device exhibits superior characteristics compared with previously reported devices (see the detailed comparison in **Supplementary Table 1**).

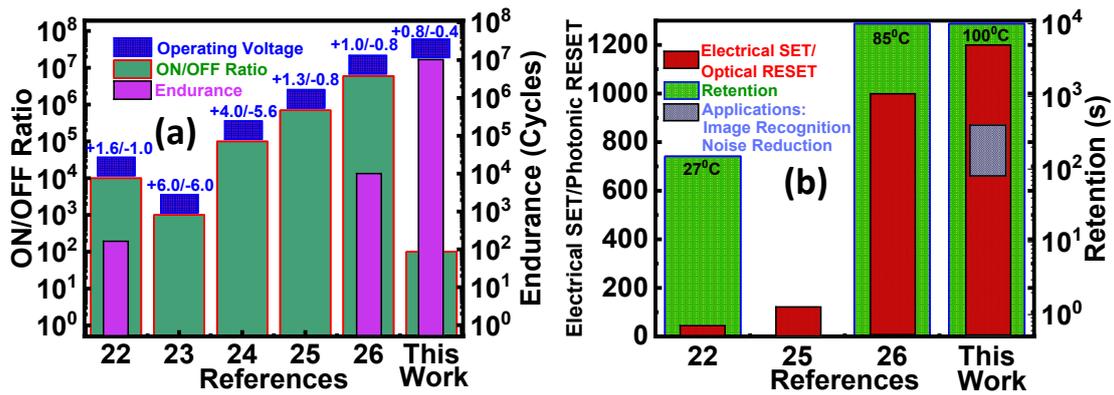

Figure 2. Benchmarking with previously reported memristors. (a) and (b) show a fair comparison of previously reported wide-bandgap oxide-based electrophotonic devices with those in the current work.

Figure 3a shows the analog DC I-V characteristic of the device with the SET voltage of approximately 0.8 V and RESET voltage of approximately –0.4 V using the compliance current of 1 mA. To check the reliability of the device, the AC endurance of the device was measured using SET voltage of 1 V and RESET voltage of –1.2 V with the speed of 200 ns, as depicted in Figure 3b. The AC durability of the device was demonstrated, highlighting its impressive ability to maintain both the low-resistance state (LRS) and high-resistance state (HRS) for over $10^7$ cycles without any signs of degradation. During the endurance tests, a read voltage of 0.2 V was used. Consistency among the different devices was determined by measuring the device-to-device variability. To this end, 20 devices were randomly selected from the wafer, and the results are shown in Figure 3c. The observed stability between the LRS and HRS indicates that the device exhibits reliable performance, making it applicable to neuromorphic computing systems. To simulate bionic synaptic plasticity, a series of voltage pulses was applied to the memristive device, as depicted in Figure 3d. Consistent LTP and LTD were achieved through the repeated application of 100 identical positive voltage pulses (1 V/200 ns), followed by 100 identical negative voltage pulses (–1.2 V/200 ns). This experimental result verifies that the synaptic weight, represented by the conductance of the device, could be increased or decreased by applying positive and negative voltage pulses, respectively. To validate the functionality of the memristive synapse, LTP and LTD were observed for different conductance (synaptic weight) states. Figure 3e and 3f show 188 LTP/LTD cycles, demonstrating the feasibility and stability of the memristive synapses over an extended period of operation. The PPF is a critical component of synaptic plasticity, and it plays a pivotal role in achieving enhanced learning and memory capabilities, as depicted in Figure 3g. PPF is achieved by applying consecutive electrical pulses and is quantified using the formula $(A2-A1)/A1 \times 100$, where A1 represents the initial excitatory current after applying the presynaptic voltage, and A2 corresponds to the increase in current observed for the second pulse, referred to as the excitatory post-synaptic current (EPSC) [36]. This increase in current following the application of a postsynaptic spike is indicative of A2. Importantly, PPF demonstrates a stable and gradual decrease in the PPF index, closely emulating the synaptic functions observed in mammalian neural systems.

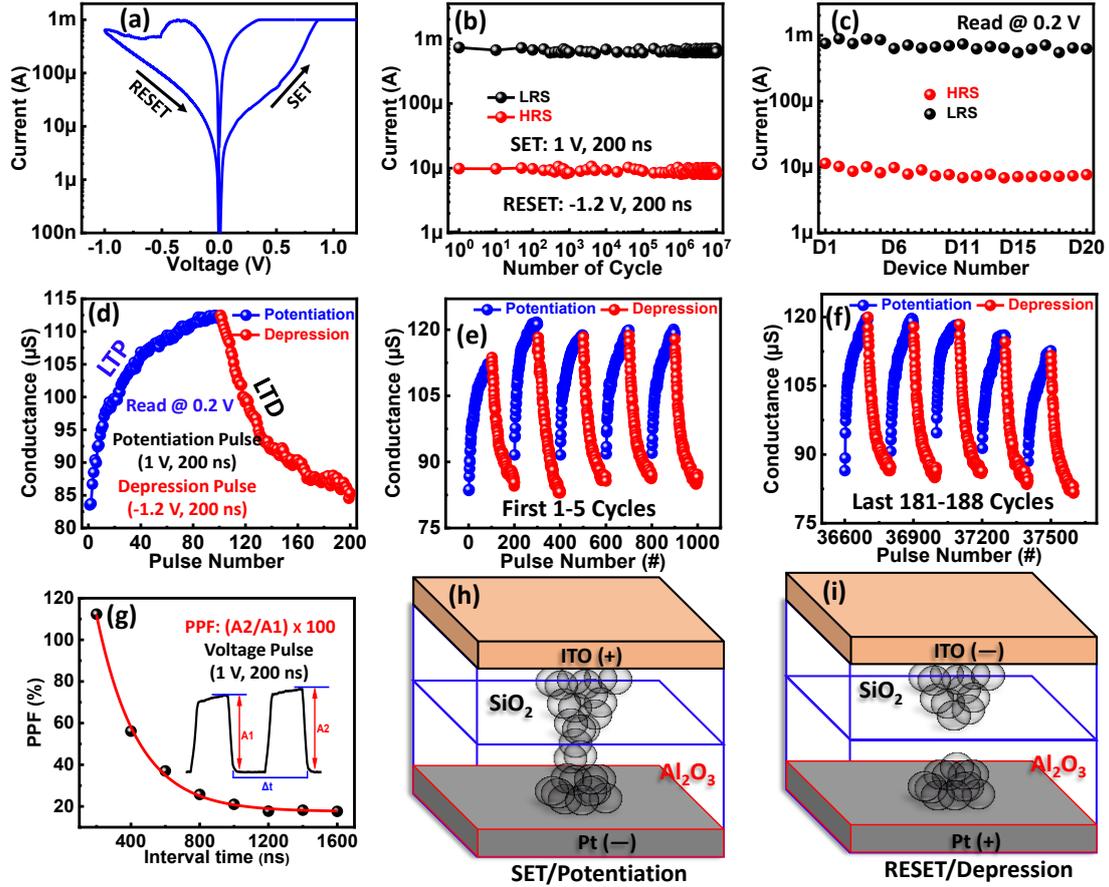

Figure 3. Electrical characterization of the memristor. a) I–V characteristic of the ITO/SiO$_2$/Al$_2$O$_3$/Pt memristor with the SET voltage of ~0.7 V and RESET voltage of ~-0.4 V. b) The AC endurance of the device with the high stability of 10$^7$ cycles in both LRS and HRS. c) The device-to-device stability of 20 randomly chosen devices from the sample. d) LTP and LTD of the device using the 100 potentiation voltage pulses and 100 depression voltage pulses. The repeatability of a) first 5 LTP/LTD cycles and b) last 5 LTP/LTD cycles with total number of 188 cycles. g) PPF of the device. h) (a) SET/Potentiation process, (d) RESET/Depression process of the ITO/SiO$_2$/Al$_2$O$_3$/Pt structure.

To confirm the excellent performance of the device, we present the conductive filament (CF) model for ITO/SiO$_2$/Al$_2$O$_3$/Pt, as depicted in Figure 3h and 3i. When a positive voltage is applied to the ITO TE, oxygen (O) vacancies are produced in the Al$_2$O$_3$ layer. These vacancies then move toward the Pt BE, leading to the formation of a cone-shaped filament within the Al$_2$O$_3$ layer. With an increase in these O vacancies, Al$_2$O$_3$ becomes electrically conductivity and works like a "virtual electrode," as described in [19]. This conductive virtual electrode, which is part of the CF structure, is primarily composed of O vacancies. Interestingly, the Al$_2$O$_3$ layer's "virtual electrode" often serves as the initiating point for the CF's regrowth within the $SiO_2$ layer. Based on the resistive switching (RS) mechanism, the observed high uniformity can be attributed to the specific rupture and recovery locations being restricted to the matrix layer close to the $SiO_2/Al_2O_3$ interface. Once the device switches to the HRS, the root of the CF does not entirely breach the RESET process. Specifically, in the ITO/$SiO_2$/$Al_2O_3$/Pt configuration, the CF tends to connect or breach primarily at the $SiO_2/Al_2O_3$ interface, and the $Al_2O_3$ layer behaves as a virtual electrode, as mentioned in [37, 38]. This RS mechanism improves the uniformity and performance of the bilayer ITO/$SiO_2$/$Al_2O_3$/Pt device.

**Density Functional Theory Calculations**

To support the conduction mechanism of the proposed ITO/$SiO_2$/$Al_2O_3$/Pt device, we used first-principles calculations based on the density functional theory and the projector-augmented-wave approach to realize the Kohn–Sham eigenvalues, electrostatic potential, and wave functions in the Vienna ab initio Simulation Package (VASP) [39]. The generalized gradient approximation (GGA) technique was utilized in the Perdew–Burke–Ernzerhof (PBE) scheme with van der Waals interactions (PBE-D2) [40-42]. The Monkhorst–Pack technique was used for the first Brillouin zone integration, and the kinetic energy cutoff was set to 500 eV [43]. A cell comprising 12 aluminum, 30 oxygen, and six silicon atoms and 6 × 6 × 1 k-point sampling were employed. The Hellmann–Feynman force operating on each atom was optimized to less than 0.01 eV/Å. As illustrated in Figure 4 (a-b), in the energy-optimized $SiO_2/Al_2O_3$ heterostructure, Al atoms formed a bond between the O (from $Al_2O_3$) and Si atoms and rearranged themselves on top of the hollow silicon dioxide region, with an interlayer spacing of 1.74 Å. The band structure is a primary determinant of the switching mechanism. We investigated the $SiO_2/Al_2O_3$ band structure with and without an applied electric field. The heterostructure exhibits metallic properties with fewer bands populated near the Fermi level, whereas in the presence of an electric field of 0.5V/nm, more bands are available near the Fermi level, making the heterostructure more conductive and easier to switch (Figure 4 c-d).

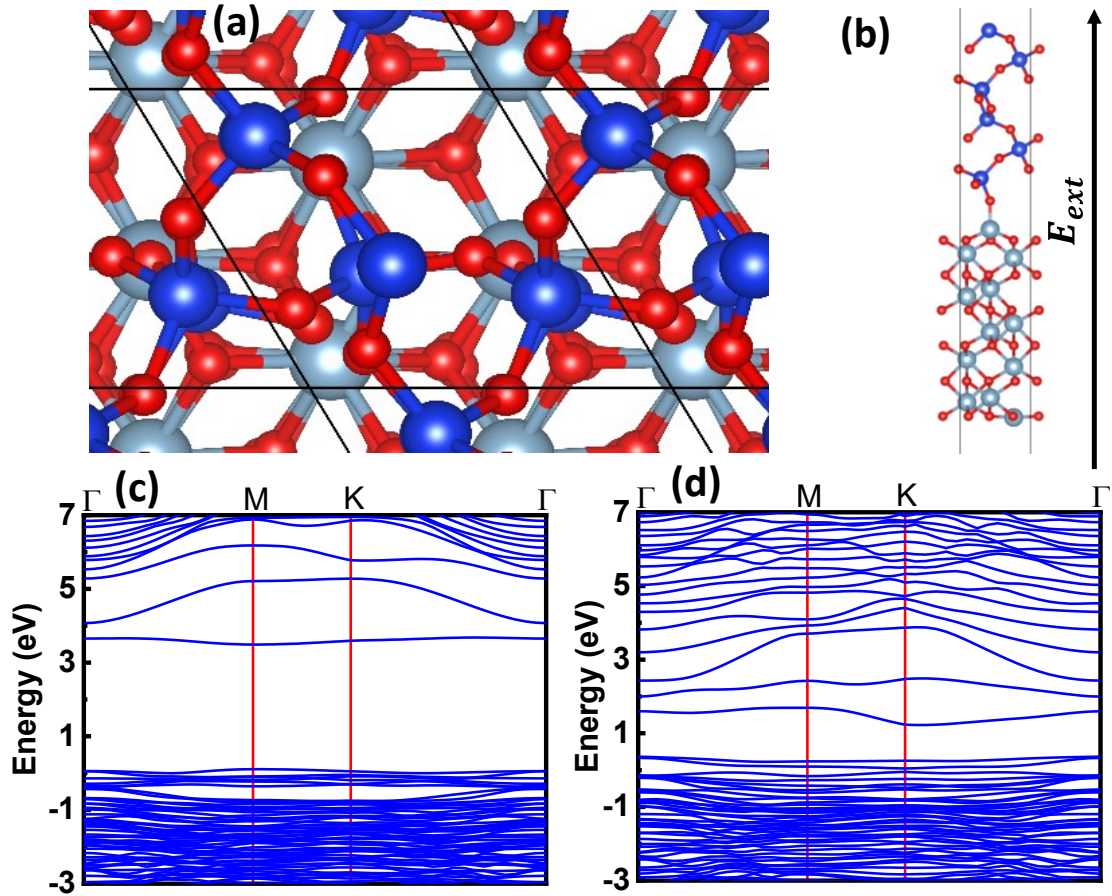

Figure 4. Density functional theory calculations. a-b) Top and side views of the $SiO_2$-$Al_2O_3$ heterostructure unit cell, the unit cell contains 12 aluminum, 30 oxygen, and six silicon atoms, the arrow shows the direction of the applied external electric field. c-d) The band structure of the $SiO_2$/$Al_2O_3$ heterostructure without and with the applied electric field of 0.5 V/nm.

**Application in Image Denoising**

Subsequently, we present a CNN autoencoder to reduce the image noise by employing an ITO/$SiO_2$/$Al_2O_3$/Pt memristor as a convolutional kernel [44, 45]. Figure 5a presents the network architecture, which comprises an encoder and a decoder. Each block includes three 3 × 3 convolutional blocks. For the training data, we manually introduced

Gaussian noise into the clean MNIST dataset images as noisy disturbances, while the original images served as training labels. Throughout training, the CNN autoencoder aims to denoise these images by eliminating noise and preserving the crucial underlying features. Following the training process, the network weight is mapped to the normalized conductance weight values, which simulate the in-memory computation with the electrical properties of the proposed device. The visualization results of the training are presented in Figure 5b, where the device was programmed to the expected value with specific pulses. We evaluated the network's efficacy using the training loss and peak signal-to-noise ratio (PSNR), as shown in Figure 5c. As the number of training epochs increased, the PSNR increased to 562, and the training loss decreased to 0.61. The experimental results demonstrate the good programming and learning capacities of the proposed $ITO/SiO_2/Al_2O_3/Pt$ device to execute in-memory computations with noise reduction applications.

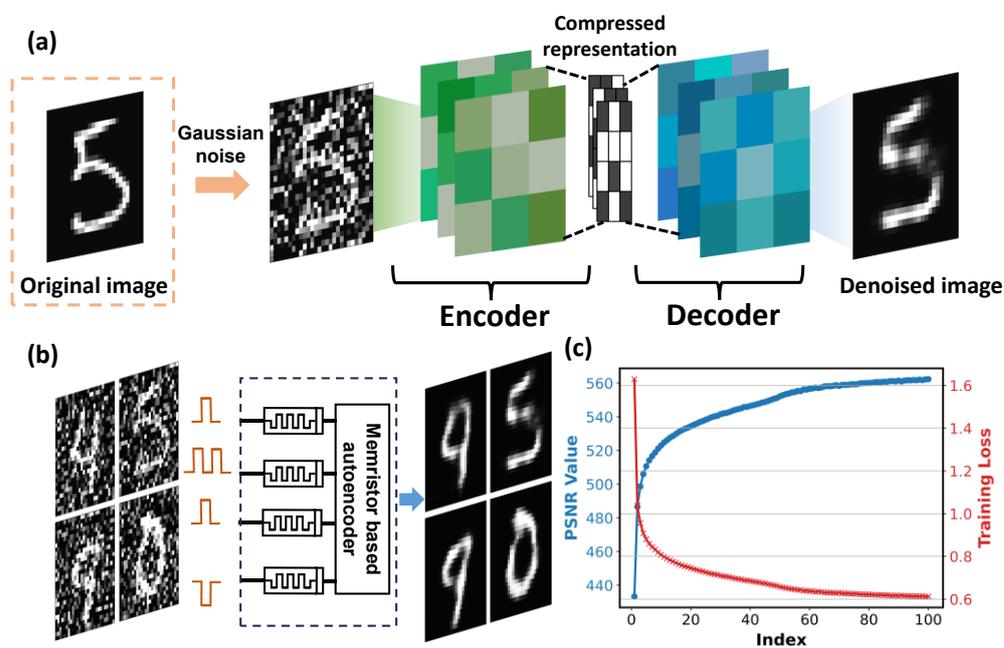

Figure 5. Image denoising application. a) Network structure of CNN autoencoder. b) Result visualization of memristor-based autoencoder. c) Training loss and PSNR result with the training epochs.

**Optoelectronic Characterization**

The electro-photonic performance and applications of the memristor in bionic synapses and the photo sensing capability of the same device are shown in Figure 6. Figure 6a shows the schematic structure of the electro-photonic memristive device under blue light (405 nm) illumination. Initially, in the dark, a positive voltage was applied to the ITO TE to change its initial state to LRS (indicated by the red curve labelled pre-illumination SET), as illustrated in Figure 6b. The formation process of the device with and without light impact is depicted in **Supplementary Figure 1**. After the pre-illumination SET, the resistance of the device was observed to be 700 μA. Following that, the blue light with intensity of 20 mW/cm$^2$ was induced on the device and the current was monitored with time at the read voltage of 0.2 V, as shown in Figure 6c. When the light was illuminated for 2 s, a sudden drop in the device's current was observed, and the LRS changed to the HRS (see the light-induced RESET process of the device with various light intensities and wavelengths in **Supplementary Figure 2**). Based on the experimental findings, we present a photo-induced interstitial ion-vacancy recombination model to elucidate the photonic RESET process. The model is derived from the widely accepted oxygen migration model initially proposed for electrically induced resistance switching in oxide-based memristor devices [18]. This aligns closely with experimental observations regarding the negative photoconductivity (NPC) phenomenon. Under the influence of an electric field, a conductive filament (CF) path forms within the oxide region owing to oxygen vacancy defects, resulting in the dissociation of metal-oxygen bonds [17, 26]. The liberated oxygen ions either migrate toward the anode with the aid of an electric field or laterally outward, facilitated by localized Joule heating at the CF site. Upon reaching the compliance limit and interrupting the applied voltage bias, the oxygen migration slows, resulting in some ions being retained in the interfacial region of the anode and the interstitial positions surrounding the CF path (see Figure 6(d)). Oxygen ions stored at the anode interface have been implicated in electrical RESET switching, a phenomenon commonly observed in bipolar memristive devices. Reversing the voltage polarity drives these ions back toward the CF path, eliminating vacancies and restoring the oxide resistance. The presence of oxygen ions at the anode interface is pivotal; a reactive metal capable of bonding with these ions at its interface ensures their subsequent release into the CF path, facilitating electrical RESET under reverse voltage polarity [17, 26]. Conversely, the interstitial oxygen ions surrounding the CF path do not contribute to the electrical RESET because they remain unaffected by the applied voltage. Nevertheless, under optical light illumination, these oxygen ions can be excited by photons, enabling them to surpass the migration barrier and migrate toward the vacancies within the CF path. Upon reaching the vacancies, recombination occurs, resulting in the annihilation of the CF path and triggering a negative photoconductivity (NPC) response in the CF current (Figure 3(e)). Ab initio simulation studies have revealed that the migration barrier of interstitial oxygen ions in common oxides is generally several hundred millielectronvolts [26, 46, 47]. As a result, it is plausible for these interstitial oxygen ions to surmount the migration barrier by absorbing photons with higher energies, generally ranging from 1.63 to 3.26 eV, within the visible light spectrum. Further investigation by Ang and Hassan [22] utilized first-principles simulations based on density functional theory. These simulations show that the electrical formation of CF within the oxide leads to the creation of oxygen vacancies, which occupy energy levels within the forbidden gap of the switching oxide. The electrical SET process is postulated to generate neutral oxygen vacancies, and the surrounding interstitial oxygen ion species are separated by an energy barrier that impedes recombination. However, charged pairs comprising oxygen vacancies and oxygen ions are prone to self-annihilation or rapid recombination via electrostatic attraction [48-50]. Light illumination positively ionizes the oxygen vacancies by exciting electrons from the trap states to the conduction band. This, in turn, induces an electrostatic attraction between the charged oxygen vacancies and interstitial oxygen ions, facilitating their recombination and subsequent disruption of the CF. To ensure the light RESET process (rupture of CF between the ITO TE and Pt BE) shown in Figure 6c, a positive voltage was applied to the ITO TE to verify the light-induced RESET, as displayed in Figure 6b. The post-illumination SET (black-labeled) validated the CF ruptures owing to light only. Figure 6f shows the repeatability of the electrical SET and light RESET processes. The device displayed at least 1200 cycles of high stability in both the LRS and HRS states without any disruption. These electro-photonic results confirm that the device is capable of photograph sensing applications. The multilevel photonic RESET and electro-photonic retention tests are depicted in **Supplementary Figure 3.**

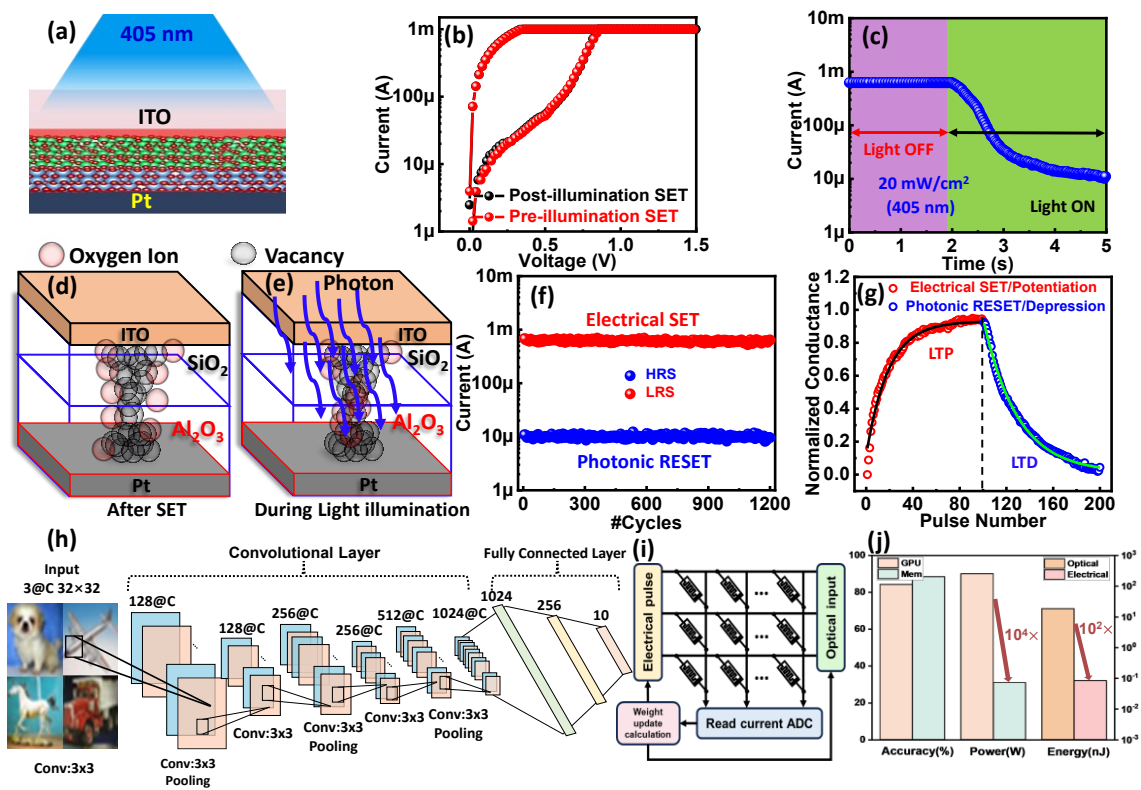

Figure 6. Optoelectronic characterization of the memristor. a) Schematic representation of the memristive

device under light illumination (405 nm). b) The electrical SET of devices before and after illuminating the light. c) Current versus time curve for the device during blue-light induced RESET using an intensity of 20 mW/cm$^2$. Three-dimensional schematic illustration of the proposed photo-assisted interstitial in-vacancy recombination model (d) Following soft electrical breakdown, a CF forms, comprising a cluster of oxygen vacancy defects. Some oxygen ions, released from the dissociated metal-oxygen bonds, disperse outward and settle at interstitial sites surrounding the conductive path once the CF process ceases. (e) Under illumination, photons stimulate these interstitial oxygen ions, prompting their migration toward the CF path. Subsequent recombination with the vacancy defects disrupts the CF path. f) Electrical SET and light RESET repeatability of the device. g) electrical LTP and photonic LTD using 100 potentiation and 100 depression pulses. h) Schematic diagram of the image recognition based on the proposed electro-photonic device. i) Circuit diagram based on electrical and photonic memristor. j) Performance comparison between a GPU and the proposed device.

In a hardware neural network utilizing SiO$_2$/Al$_2$O$_3$ electro-photonic synaptic devices, we assigned weights to the multilevel conductance states (as depicted in Figure 6g). During training, weight updates were controlled by both electrical and photonic pulses. Specifically, 100 electrical pulses with an amplitude of 1 V and a pulse width of 200 ns (read voltage: 0.2 V) are employed for LTP, whereas 100 photonic pulses with a wavelength of 405 nm and a pulse width of 0.01 s (read voltage: 0.2 V) are used for LTD. An ideal synaptic device exhibits a linear weight update that is directly proportional to the number of electrical/optical pulses applied for potentiation/depression. Nevertheless, the proposed SiO$_2$/Al$_2$O$_3$ electro-photonic synaptic devices exhibit nonlinearity in both the LTP and LTD characteristics, causing deviations from the ideal linear weight change [51]. Consequently, the nonlinearity in the changes to the conductance significantly affects the accuracy of the neural network during the training process. To quantify the impact of this nonlinearity, a behavioral model was employed to evaluate the linearity of the weight updates using the following equations [52]:

$$G_{LTP} = B\left(1 - e^{-\frac{P}{A}}\right) + G_{min} \tag{1}$$

$$G_{LTD} = -B\left(1 - e^{\frac{P-P_{max}}{A}}\right) + G_{max} \qquad (2)$$

$$B = \frac{G_{max} - G_{min}}{1 - e^{-\frac{P_{max}}{A}}}, \qquad (3)$$

where $G_{LTP}$ and $G_{LTD}$ represent the conductance values for LTP and LTD, respectively. $G_{max}$, $G_{min}$, P, and $P_{max}$ correspond to the maximum and minimum conductance, pulse number, and maximum pulse number, respectively, all of which were directly derived from the experimental data (Figure 6b). A is the nonlinearity parameter that governs the nonlinear behavior of the weight update. For the given experimental data, we calculated nonlinearity values of 2.8 for LTP and 1.8 for LTD, reflecting the degree of nonlinearity in these processes.

To demonstrate the versatility of the proposed $SiO_2/Al_2O_3$ electro-photonic synaptic devices in the context of artificial intelligence hardware, we applied them to the classification of the Canadian Institute for Advanced Research (CIFAR-10) dataset with a nine-layer CNN, which is a simplified variant of the VGG-NET [53, 54], as illustrated in Figure 6h. Since the adoption of the CNN model for these devices, unprecedented success has been achieved in tackling computer vision challenges, including image segmentation and object detection, surpassing the performance of traditional neural networks. Our network architecture comprises six convolutional layers for feature extraction and three fully connected layers for image classification. After every two convolutional layers, we adopt a max-pooling layer behind the subsample and aggregate the features. In our simulation, the synaptic values of our device were derived and mapped to the trained CNN weights (please see **Supplementary Figures 4 and 5** for a detailed explanation). Figure 6i shows an ideal circuit diagram of the memristor array with both electrical and optical responses. The current flowing through the array can be accumulated naturally through Kirchhoff's law, which helps accelerate matrix computation. The weight-update module aligns with the input pulses and adjusts the conductance change based on the feedback. Figure 6j presents a comparison of the training results and power consumption. The ideal training result for the GPU achieves a recognition accuracy of 88.43%, whereas the memristor-based simulation exhibits a marginally lower accuracy of 84.23%. However, in terms of energy efficiency, we evaluated the power consumption using traditional GPU and memristor arrays [55]. The proposed device demonstrates a significant advantage in that it consumes $10^4$ times less energy than a Tesla P40 GPU. Moreover, we compared the energy cost per pulse between the noteoptical and electrical programming methods, revealing that the optical-based method reduced the energy consumption by $10^2$ units. The evaluation of the power and energy costs is explained in **Supporting Information Note-7**.

In conclusion, in this study, we engineered a CMOS-compatible ITO/SiO$_2$/Al$_2$O$_3$/Pt memristor device to demonstrate its excellent capabilities in terms of LTP and LTD, exhibiting strength through a minimum of 188 repetitive LPT/LTD cycles with a total of 37600 pulses. Endurance is pivotal to ensure the reliability and synaptic strength of a device. Furthermore, we assessed the PPF, a crucial biological feature of synaptic devices, by applying successive AC pulses. In addition to its role as a synaptic element, the device was seamlessly integrated into a CNN autoencoder to reduce image noise. First-principles calculations showed that the SiO$_2$/Al$_2$O$_3$ heterostructure becomes more conductive when an electric field is present because the bands become more metallic, which helps filament formation in the resistive switching layers. Notably, the proposed device exhibited outstanding electrical SET and light RESET endurance, maintaining stability for a minimum of 1,200 cycles without any performance degradation. It showcases good data retention capabilities, persisting for up to $10^4$ seconds even when exposed to a temperature of 100 °C. We conducted experiments with the device using an autoencoder and a CNN, specifically for noise reduction and image recognition applications, achieving a PSNR of 562 and an accuracy rate of 84.23%. This pioneering research not only paves the way for the incorporation of our device into image processing applications but also signifies a significant advancement in the field of in-memory computing and photo-sensing capabilities within a single cell.

**Experimental Section**

*Device Fabrication*: The proposed device was fabricated on a 4-inch Si wafer. Before fabrication, the Si wafer was cleaned with isopropyl alcohol (IPA) and deionized (DI) water and then dried using $N_2$ gas. First, a 250 nm $SiO_2$ was grown on Si wafer by plasma-enhanced chemical vapor deposition (PECVD) at 350° C. Subsequently, the bottom electrode was patterned by optical lithography on 4-inch silicon wafer. After Pt, 5 nm $Al_2O_3$ and 5 nm $SiO_2$ as switching layers were grown via atomic layer deposition (ALD) at 250 °C and PECVD at 350 °C, respectively. Subsequently, a 100 nm ITO top electrode was deposited via sputtering and patterned through a liftoff process to prepare an ITO/$SiO_2$/$Al_2O_3$/Pt memristive crossbar array with a cell size of 20 × 20 $\mu m^2$ (**Please see the Supplementary Figure 6 for details fabrication process of crossbar array**).

*Characterization and Measurement:* The 'electrical characteristics of the device were measured using a Keysight B1500A semiconductor device parameter analyzer. For the photoinduced measurements, a visible blue light-emitting diode (LED) source (405 nm, Shanghai Dream Lasers Technology) and an electronic shutter controller (Newport) were employed.

**Acknowledgements:** Authors acknowledge the generous financial support provided by the Semiconductor Initiative at King Abdullah University of Science and Technology. More specifically, this publication is based on work supported by King Abdullah University of Science and Technology (KAUST) Research Funding (KRF) under

Award No. ORA-2022-5314. HL acknowledges the support of the National Research Foundation of Korea (NRF) Grant (No. 2021R1A5A103299611) funded by the Ministry of Science, ICT, and Future Planning (MSIP) of the Korean government.
**Authors Information:**

Dr. Dayanand Kumar, Mr. Hanrui Li, Mr. Manoj Kumar Rajbhar, Mr. Abdul Momin Syed, and Prof. Nazek El-Atab* are with the Smart Advanced Memory Devices and Applications (SAMA) Laboratory, Electrical and Computer Engineering, King Abdullah University of Science and Technology (KAUST), Thuwal 23955-6900, Saudi Arabia. Corresponding email address: Nazek.elatab@kaust.edu.sa

Mr. Amit Kumar is with the Department of Physics and Astronomy, University of Manchester, Manchester M13 9PL, UK.

Prof. Hoonkyung Lee is with the Department of Physics, Konkuk University, Seoul 05029, Republic of Korea.

**Authors Contribution:**

N.E.A conceived the idea and supervised the work. D.K. and M.K.R. fabricated the devices. D.K. characterized the devices. A.S. performed SEM imaging. H.L. performed the neural network simulations. A.K. conducted the DFT calculations under H.L. supervision. All authors contributed to writing the manuscript.

**Competing Interests:**
The authors declare that there are no competing interests.

**Data Availability:**
The datasets generated during and/or analyzed during the current study are available from the corresponding author on reasonable request.